\documentclass[aip,article]{revtex4-1}

\def\bk{\bf k}
\newcommand{\be}{\begin{equation}}
\newcommand{\ee}{\end{equation}}
\newcommand{\beq}{\begin{eqnarray}}
\newcommand{\eeq}{\end{eqnarray}}

\usepackage{amsmath}
\usepackage{graphicx}

\begin{document}

\title{Casimir force induced by imperfect Bose gas}
\author{Marek Napi\'orkowski}
\email{Marek.Napiorkowski@fuw.edu.pl} 
\author{Jaros{\l}aw Piasecki}
\affiliation{Institute of Theoretical Physics, Faculty of Physics,\\ University of Warsaw\\
 Ho\.za 69, 00-681 Warsaw, Poland} 

\begin{abstract}
{We present a study of the Casimir effect
in an imperfect (mean-field) Bose gas contained between two infinite parallel 
plane walls. 
The derivation of the Casimir force follows from the calculation of the excess grand canonical free energy density 
under periodic, Dirichlet, and Neumann boundary conditions with the use of the steepest descent method.
In the one-phase region the force decays exponentially fast when distance $D$ between the walls
tends to infinity. 
When Bose-Einstein condensation point is approached the decay length in the exponential law
diverges with critical exponent  $\nu_{IMP}=1$, which differs from the perfect gas case where $\nu_{P}=1/2$.
In the two-phase region the Casimir force is long-range, and decays following
the power law $D^{-3}$, with the same amplitude as in the perfect gas. }
\end{abstract}
\maketitle

\section{Introduction}

We present a study of the Casimir effect \cite{GAM2009, GAM2011, KR1994, SY1981, KD1992, KG1999, RP2005} 
occuring when an \emph{imperfect} Bose gas fills a
region between  two  infinite parallel plane walls separated by distance $D$. In particular, we are interested 
in the properties of the fluctuation-induced force between the walls as function of  distance $D$, and of the 
thermodynamic state of the imperfect Bose gas in the vicinity of the Bose-Einstein condensation point. Our work 
is thus a contribution to the theory of critical Casimir
effects in condensed matter physics.\\

The case of a \emph{perfect} Bose gas has been analyzed by Martin and Zagrebnov \cite{MZ2006}.
The authors considered the grand canonical potential $\Omega (T,L,D,\mu)$ of a perfect Bose gas filling
a box (rectangular parallelepiped) with linear dimensions $L\times L\times D$. The temperature and the 
chemical potential are denoted by $T$ and $\mu$, respectively. Periodic, Dirichlet, and Neumann 
boundary conditions were considered. The main result of their study was the 
determination of the large $D$ asymptotics of the grand canonical potential per unit wall area
\be
\label{gce}
\omega (T,D,\mu ) = \lim_{L\to\infty} \frac{\Omega (T,L,D,\mu)}{L^2} 
\ee
It turns out that Bose-Einstein condensation induces a qualitative change in the 
behavior of the excess grand canonical free energy $\omega_{s}(T ,D,\mu)$ 
\be
\label{correction}
\omega_{s}(T ,D,\mu) =   [ \omega (T,D,\mu) - D\,\omega_{b}(T,\mu)  ] \quad, 
\ee
where $\omega_{b}$ denotes the potential $\Omega$ per unit volume evaluated in the thermodynamic limit. The excess grand canonical 
free energy $\omega_{s}(T ,D,\mu)$ contains contributions due to each single  wall confining the system and the interaction between the confining walls. 

In the one-phase region corresponding to $\mu < 0$,  the force 
\be
\label{FD}
F(T,D,\mu) = -\frac{\partial \omega_{s}(T,D,\mu)}{\partial D}
\ee
decays exponentially fast for $D\to\infty$, whereas in the presence of condensate, when $\mu =0$, there appears a long-range 
{\it Casimir force} 
\be
\label{CasimirF}
\frac{F(T,D,0)}{k_{B}T} =  \frac{\Delta}{D^3} \quad.
\ee
The Casimir force as defined in Eq.(\ref{FD}) is evaluated per unit area and has dimension of pressure. The amplitude $\Delta$ depends 
on the type of boundary conditions imposed at the walls and is otherwise universal. 
For example, in the case of periodic boundary conditions the amplitude $\Delta$ is negative and corresponds to attractive Casimir forces. 
The analysis presented in \cite{MZ2006} was extended in \cite{GD2006}, where the dependence of the Casimir force on thickness $D$ and 
temperature was presented with the help of universal scaling function of variable $D/\xi$, where $\xi$  is the bulk correlation length of density fluctuations. 
This extension contains both the Casimir effect at criticality and its asymptotic behavior off criticality as special cases. 

Our object in the present study is to generalize the above analysis to the case of an imperfect (mean-field)
Bose gas. The repulsive pair interaction between identical bosons is described in this model by associating
with each pair of particles the same mean energy $(a/V)$, where $a>0$, and $V$ denotes the volume
occupied by the system. The Hamiltonian of the imperfect Bose gas \cite{D1972} composed of $N$ particles
\be
\label{HMF}
H  = H_{0} + H_{mf} 
\ee 
is the  sum of the kinetic energy 
\be
\label{HPG}
H_{0} = \sum_{\bk}\frac{\hbar^2 \bk^{2}}{2m}{\hat n}_{\bk} \quad,
\ee 
and the term representing the mean-field approximation to the interparticle interaction
\be
\label{mf}
H_{mf} = \frac{a }{V}\frac{N^2}{2 } \quad. 
\ee 
The symbols $\{{\hat n}_{\bk}\}$  denote the particle number operators and the summation is over one-particle states $\{\bk\}$. 
We follow here the generally adopted definition of the {\it imperfect
Bose gas} where the exact number of pairs $N(N-1)/2$ is replaced in
$H_{mf}$ by $N^2/2$ (see the corresponding comment in \cite{ZB2001}).
  As far as the applicability of the mean-field approximation is
concerned, an important rigorous result is to be noted. It turns out that
the mean field theory can be equivalently formulated with the help of a
repulsive binary potential subject to the so-called Kac's scaling 
$ V_{\gamma}(r) = \gamma^{3}v(\gamma r) $, where $v(r)$ is the actual binary potential.  
When the dimensionless parameter $\gamma$  approaches zero, the potential
$V_{\gamma}(r)$ becomes weak and long range, leading exactly to the mean
field theory based on the Hamiltonian in Eq.(\ref{HMF}) with constant $a$ equal to the
$\gamma$-independent space integral of the potential $V_{\gamma}$ \cite{L1986}. This remarkable 
connection clarifies the physical content of the mean field model studied here. 

The thermodynamics of the imperfect Bose gas has been rigorously studied showing that Bose-Einstein
condensation persists in the presence of mean-field interaction \cite{BLS1984}. 
The corresponding critical density $n_{c}$ and critical temperature $T_{c}$ are the same as in 
the perfect gas case. Only the critical chemical potential is shifted from $\mu=0$ in the perfect gas case 
to a positive value $\mu_{c}=a n_{c}$. 
We can thus expect the appearance of Casimir forces much as in the case of the perfect gas. 
However, the imperfect Bose gas is a qualitatively different system which motivates the present study. 
Indeed, the presence of the term $H_{mf}$ makes the system's Hamiltonian superstable which, in turn, 
implies equivalence between canonical and grand-canonical ensembles \cite{D1972,ZB2001}. This is not the case
for the perfect gas where the grand-canonical description leads to unphysical macroscopic fluctuations
in the overall density in the ground state, whereas the canonical ensemble predicts physically satisfactory
behavior \cite{ZUK77, BP1983}. Moreover, the thermodynamics of the imperfect Bose gas is defined for all values of
the chemical potential, whereas the states of an ideal Bose gas are confined to the region $\mu \le 0$.
In the present paper we thus investigate Casimir forces in physically more satisfactory case of 
interacting mean-field Bose gas. 

In Section II we present an original derivation of the bulk grand-canonical free energy density $\omega_{b}(T,\mu)$ using the steepest descent method. 
The same method is then applied to determine the Casimir force and study its scaling properties in Section III. The last Section IV contains concluding comments.

\section{Bulk properties of the imperfect Bose gas via the steepest descent method}

We consider the imperfect Bose gas composed of $N$ identical particles of mass $m$ contained in $L \times L \times D$
rectangular parallelepiped of volume $V=L^{2}D$; $D$ is the distance between two $ L \times L $ walls.

Owing to the structure of the Hamiltonian (\ref{HMF}) the grand canonical partition function $\Xi(T,L,D,\mu)$ can be conveniently written in the 
occupation numbers representation  
\begin{eqnarray}
\label{part1}
\Xi(T,L,D,\mu)\,=\, \sum_{N=0}^{\infty} \exp(\beta \mu N) \,{\sum}_{\{n_{\bk}\}}^{'} \,\exp\left[-\beta \left(\sum_{\bk} n_{\bk} \epsilon_{k} + 
\frac{a}{2}\frac{N^2}{V} \right)\right] \quad,
\end{eqnarray}
where $\sum_{\{n_{\bk}\}}' $ denotes summation over all sets $\{n_{\bk}\}$  of occupation numbers of one-particle states 
$\{\bk\}$ under the constraint $ N=\sum_{\bk} n_{\bk}$. The kinetic energy in state $\bk$ equals 
$\epsilon_{\bk} = \hbar^2|{\bk }|^2 /2m = \hbar^2 k^2/2m $, and  $\beta = 1/k_{B}T$.  

The partition function can be rewritten as
\begin{eqnarray}
\label{part2}
\Xi(T,L,D,\mu)\,=\,\exp\left(\frac{\beta V}{2a}\,\mu^2\right)\,\sum_{N=0}^{\infty} \, \exp\left[-\frac{\beta a}{2V}\,(N-\frac{V\mu}{a})^2\right] Z_{0}(T,L,D,N) \quad,
\end{eqnarray}
where $Z_{0}(T,L,D,N)$ denotes the canonical partition function of a pefect Bose gas 
\begin{eqnarray}
 Z_{0}(T,L,D,N) =  {\sum}^{'}_{\{n_{\bk}\}} \,\exp\left(-\beta \sum_{\bk} n_{\bk} \epsilon_{k} \right) \quad.
\end{eqnarray}
It is useful to rewrite Eq.(\ref{part2}) by introducing an arbitrary real parameter $\alpha$  in the following way 
\begin{eqnarray}
\label{part3}
\Xi(T,L,D,\mu)\,= \,\exp\left[\frac{\beta V}{2a}\,(\mu - \alpha)^2\right]
\end{eqnarray}
\[ \times\sum_{N=0}^{\infty} \, 
 \exp\left\{-\frac{\beta a}{2V}\,\left[N-\frac{V}{a}\,(\mu - \alpha)\right]^2\right\} \, \exp(\beta N \alpha ) \, Z_{0}(T,L,D,N) \quad,   \]
where the rhs of Eq.(\ref{part3}) actually does not depend on $\alpha$. Such a representation of  $\Xi(T,L,D,\mu)$ has been already used in the 
proof of condensation taking place in the imperfect Bose gas \cite{BLS1984}. 

The identity 
\begin{eqnarray}
\exp\left(-\frac{\gamma^2}{4\delta}\right) \,=\, \sqrt{\frac{\delta}{\pi}} \, \int_{-\infty}^{\infty} dq \, 
\exp(-\delta q^2 + i \gamma q)
\end{eqnarray}
valid for arbitrary $\delta > 0$, permits one to establish relation between $\Xi(T,L,D,\mu)$ and the grand partition function
of a perfect gas $\Xi_{0}(T,L,D,\mu)$ 
\be
\label{Xiideal}
\Xi_{0}(T,L,D,\mu) =\, \sum_{N=0}^{\infty} \exp(\beta \mu N) \,Z_{0}(T,L,D,N)
\ee 
evaluated for a complex value of the chemical potential $(\alpha + iq)$. Indeed, the following sequence
of equalities holds

\begin{eqnarray}
\label{part4}
& \Xi&(T,L,D,\mu) =  \exp\left[\frac{\beta V}{2a}\,(\mu - \alpha)^2 \right] \\
&\times&  \sum_{N=0}^{\infty} \sqrt{\frac{V\beta}{2\pi a}}\,
\int_{-\infty}^{\infty} dq \,\exp\left[- \frac{V\beta}{2a}q^2 + iq\beta\,[N-\frac{V}{a}\,(\mu - \alpha)] +\beta \alpha N \right] 
Z_{0}(T,L,D,N)  \nonumber
\end{eqnarray}
\[ =  \,\exp\left[ \frac{\beta V}{2a}\,\mu^2 \right] \sqrt{\frac{V \beta}{2\pi a}}\,\int_{-\infty}^{\infty} dq \,\exp\left[ \frac{V\beta}{2a}(iq+\alpha)^2 - 
(iq+\alpha)\frac{V \beta \mu}{a}\right]  \, \Xi_{0}(T,L,D,\alpha + iq) \]
\[ =  (-i)\,\exp\left[ \frac{\beta V}{2a}\,\mu^2 \right] \sqrt{\frac{V \beta}{2\pi a}} \int_{ \alpha - i\infty}^{\alpha + i\infty } dt 
\,\exp\left[ \frac{\beta V}{a}\,\left(\frac{t^2}{2}-t\mu \right)\right]\,\Xi_{0}(T,L,D,t) \quad \]\\[0.01cm]

Relation (\ref{part4}) between $\Xi(T,L,D,\mu)$ and $\Xi_{0}(T,L,D,t)$ will be the basis for subsequent analysis.
The parameter $\alpha$ appears here only in the contour of integration. Due to its arbitrariness we can choose $\alpha <0$ to ensure 
the convergence of the series defining the analytic continuation of the perfect gas partition function (\ref{Xiideal}).

We will apply here the method of steepest descent (saddle point method) to determine the large $L$ and large $D$ asymptotics of the
partition function $\Xi(T,L,D,\mu)$. This, in turn, will yield the bulk free energy density (per unit volume) as well as the 
excess grand canonical free energy density (per unit area). 

To illustrate the method we begin by evaluating in this section the bulk free energy density. 
This is conveniently done by considering a cubic volume $V=L^3$. We choose the coordinate system with axes  oriented along the edges of the 
cube adopting in each direction periodic boundary conditions.  The rhs of Eq.(\ref{part4}) can be then rewritten in the following way
\begin{eqnarray}
\label{part5}
\Xi(T,L,L,\mu)\,=\,-i\,\exp\left[{\frac{\beta V}{2a}\,\mu^2}\right]\,\sqrt{\frac{V}{2\pi \beta a}}\,\int_{\beta \alpha - i\infty}^{\beta \alpha + i\infty} ds 
\,\exp[-V\,\varphi_{b} (s)] \quad,
\end{eqnarray}
where
\begin{eqnarray}
\label{part6}
\varphi_{b} (s)\,=\, \frac{1}{\beta a}\,\left(-\frac{s^2}{2}\,+\,s \beta \mu \right) - \frac{1}{\lambda^3}\,g_{5/2}[\exp(s)]\,+\,\frac{1}{V}\,\ln[1-\exp(s)] \quad.
\end{eqnarray}
Here $\lambda = h/\sqrt{2 \pi m k_B T}$ denotes the thermal de Broglie wavelength , and $g_{\kappa}(z) $ is the Bose function 
\begin{eqnarray} 
g_{\kappa}(z) = \sum_{k=1}^{\infty} \frac{z^k}{k^\kappa} \quad. 
\end{eqnarray}
Equation (\ref{part6}) follows directly from the formula 
\be
\label{perfect}
\frac{1}{V} \ln\Xi_{0} (T,L,L,\mu) = \frac{1}{\lambda^3} g_{5/2}[\exp(\beta \mu)] - \frac{1}{V}\ln[1-\exp(\beta \mu)]
\ee
valid for $V/\lambda^3 \gg 1$ (see e.g.  \cite{ZUK77, ZB2001}). It is well known that the theory of Bose-Einstein condensation in a perfect gas 
requires separate examination of the contribution $[- \ln[1-\exp(s)]/V ]$ stemming from the 
one-particle ground state. 

In the steepest descent method \cite{ST1973} one has to find the saddle point $s_0$ at which the derivative of the 
analytic function $\varphi_{b}(s)$ vanishes
\begin{eqnarray}
\label{root1} 
0=\varphi_{b}'(s_0)= \frac{1}{\beta a}\,(-s_0 + \beta \mu) - \frac{1}{\lambda^3}\,g_{3/2}[\exp(s_{0})]-\frac{1}{V}\,
\frac{\exp(s_{0})}{1-\exp(s_{0})} \quad.
\end{eqnarray}
This equation can be rewritten with the help the critical density $n_c$ for the perfect Bose gas
\begin{eqnarray}
\lambda^3 \, n_c \,=\, g_{3/2}(1) = \zeta(3/2) = 2.612... \quad
\end{eqnarray}
in the following way
\begin{eqnarray}
\label{root11}
-\frac{s_{0}}{\beta a n_{c}} \,+ \, \frac{\mu}{a\,n_{c}}\,=\,\frac{g_{3/2}[\exp(s_{0})]}{g_{3/2}(1)}\,+\frac{1}{V\,n_{c}}\,
\left[ \frac{\exp(s_{0})}{1-\exp(s_{0})}\right] \quad.
\end{eqnarray}
Depending on the value of parameter $(\mu /a\,n_{c})$ one has to consider separately two cases:  
$ (\mu /a\,n_{c})\,< 1 $, and $(\mu /a\,n_{c})\,> 1 $. \\
In the first case, when  $(\mu /a\,n_{c})\,< 1  $, there exists a unique real negative solution ($s_0 < 0$) satisfying the equation
\begin{eqnarray}
\label{root2}
-\frac{s_{0}}{\beta a n_{c}} \,+ \, \frac{\mu}{a\,n_{c}}\,=\,\frac{g_{3/2}[\exp(s_{0})]}{g_{3/2}(1)} \quad. 
\end{eqnarray}
This follows directly from the analysis of the point of intersection of the straight line
\be
\label{line}
f(s_0) = -\frac{s_{0}}{\beta a n_{c}} \,+ \, \frac{\mu}{a\,n_{c}}
\ee
with the curve
\be
\label{curve}
g(s_0) = \frac{g_{3/2}[\exp(s_{0})]}{g_{3/2}(1)}
\ee
represented in Fig.1. 
\begin{figure}[hbt]
\includegraphics{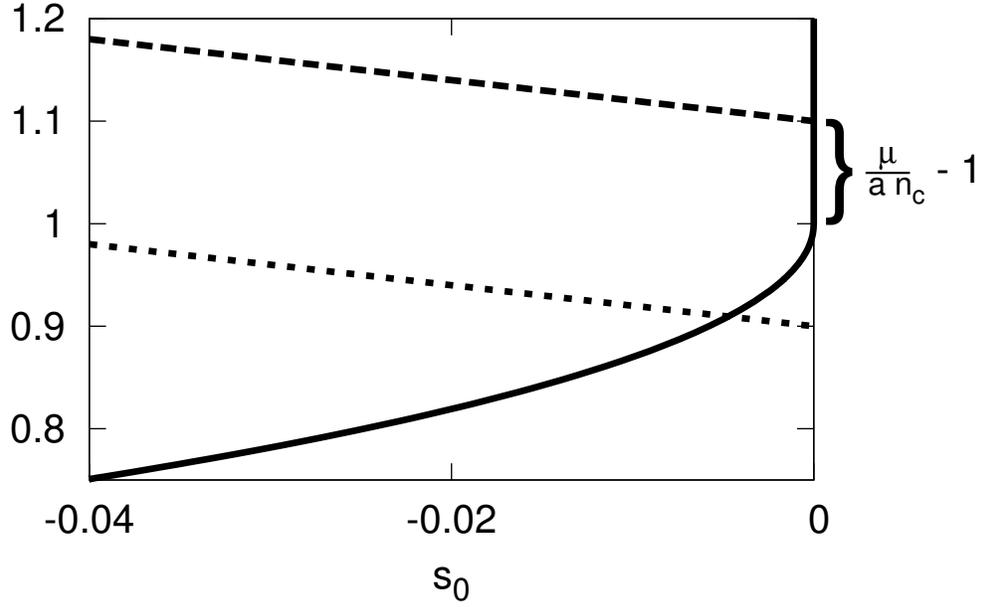}
 \begin{center}
  \caption{Graphical solution of Eq.(\ref{root11}). Two straight lines represent the lhs of Eq.(\ref{root2}) in two cases: 
	the dotted line corresponds to $\frac{\mu}{a\,n_{c}} =0.9$ and $\beta a n_{c} =0.5$ (one-phase region); the dashed 
	line corresponds to $\frac{\mu}{a\,n_{c}} = 1.1$ and $\beta a n_{c} =0.5$ (two-phase region). The continuous line 
	corresponds to the rhs of Eq.(\ref{root11}) in the limit $V \rightarrow \infty$.  The density of the condensate $\mu/a-n_{c}$ is proportional to 
	the segment denoted by the bracket on vertical axes.}
 \end{center}
\end{figure}
Clearly, when $s_0 < 0$, the second term on the rhs of (\ref{root11}) does not contribute in the bulk limit $V\to\infty$.
We also note that the saddle point $s_{0}(T,\mu)$ tends to zero from below when approaching the imperfect gas condensation 
point $\mu \, =\, a n_{c}$. 

Applying the method of steepest descent \cite{ST1973} to determine the large $V$ asymptotics of
 the integral in Eq.(\ref{part5}) we find that the bulk free energy density 
\begin{eqnarray}
\omega_{b}(T,\mu) = - \lim_{L \rightarrow \infty} \frac{1}{L^3}\,k_{B}T\,\ln \Xi(T,L,L,\mu)\,= -p 
\end{eqnarray}
is given by 
\begin{eqnarray}
\label{bulkdensity}
\omega_{b}^{<}(T,\mu)= -\frac{\mu^2}{2a} - \frac{1}{\beta^2 a}\,\left(\frac{s_{0}^2}{2}-s_{0} \beta \mu \right) - 
\frac{1}{\beta \lambda^3}\,g_{5/2}[\exp(s_{0})] \quad,
\end{eqnarray}
where $p$ denotes the pressure and the superscript ${}^{<}$ indicates that we consider the case of $\mu \, < \, a n_{c}$.  

The bulk density $\omega_{b}^{<}(T,\mu)$ can be equivalently rewritten  as
\begin{eqnarray}
\label{free1}
\omega_{b}^{<}(T,\mu)= -p^{<}(T,\mu) = -\frac{a}{2} \, x^2(T,\mu) \,-\, \frac{1}{\beta \lambda^3}\,g_{5/2}[\exp (\beta (\mu - a x(T,\mu))] \quad,
\end{eqnarray}
where $x(T,\mu)$ satisfies the equation
\begin{eqnarray}
\label{free11}
x = \frac{1}{\lambda^3}\,g_{3/2}[\exp[\beta (\mu - a x)] \quad.
\end{eqnarray}
In fact, a straightforward calculation shows that
\be
\label{density}
\frac{\partial}{\partial \mu}\omega_{b}^{<}(T,\mu) = - x(T,\mu) 
\ee
so that $x(T,\mu) = n(T,\mu)$  is the number density of the gas expressed as function of
chemical potential and temperature.

In the complementary case  $ (\mu /a\,n_{c})\,> 1$ the solution of Eq.(\ref{root11}) is obtained by taking the limit
$s_{0}\to 0, L \to \infty$ , and requiring the equality
\be
\label{condensate}
 \frac{\mu}{a}\,=\, \,n_{c} + \lim_{s_{0}\to 0, L\to\infty}\frac{1}{L^3\,}\,\left[ \frac{\exp(s_{0})}{1-\exp(s_{0})}\right] 
\ee
\[  = n_{c} +  \lim_{s_{0}\to 0, L\to\infty}\frac{1}{L^3\,}\,\left[ \frac{1}{1-\exp(s_{0})}\right] 
  = n_{c} + (\rm{ density\; of \; condensate})   \quad,  \] 
	see the dashed line in Fig.1. 
The precise meaning of the double limit $s_{0}\to 0, L\to\infty$ defining the density of condensate in the perfect gas
theory can be found in \cite{ZUK77, ZB2001}.

The bulk free energy density in the region $(\mu /a\,n_{c})\,> 1$ is thus obtained by taking the limit  
$s_{0}\to 0$ in Eq.(\ref{bulkdensity}) , so that  
\begin{eqnarray}
\label{free2}
\omega_{b}^{>}(T,\mu)= -\frac{\mu^2}{2a} \,-\, \frac{1}{\beta \lambda^3}\,\zeta(5/2) \quad. 
\end{eqnarray}
or
\be
\label{equstate}
p^{>}(T,\mu) = \frac{\mu^2}{2a} + p_{0}(T,\mu =0)
\ee
where $p_{0}(T,\mu =0) $ is the perfect gas pressure in the presence of condensate.  
Eq.(\ref{equstate}) implies the relation $n = \mu/a$, valid in the presence of condensate.

The above results for the bulk density of the grand canonical potential 
yield equations of state (\ref{free1}),(\ref{free11}),(\ref{equstate}),  in full agreement with  rigorous results derived  by more complex methods
(see e.g. \cite{PZ2004}). 

\section{Casimir force between parallel plane walls}

In order to evaluate the Casimir force (per unit area) acting between infinite parallel plane walls separated 
by distance $D$ one has to determine the corresponding excess grand canonical free energy density (per unit area) defined by
\begin{eqnarray}
\label{root3}
\omega_{s}(T,D,\mu) = \lim_{L\to\infty} \left[ \frac{\Omega(T,L,D,\mu)}{L^2} \right] - D\,\omega_{b}(T,\mu) \quad.
\end{eqnarray}
The system surface properties, and the excess grand canonical free energy in particular, depend on the boundary conditions imposed 
on the confining walls. Here we concentrate on the case of periodic boundary conditions although the generalization 
to the Dirichlet and Neumann cases is straightforward. We choose the coordinate system whose $z$-axis is perpendicular to 
the confining walls.

It is convenient to rewrite the partition function (\ref{part4}) in the following way 
\begin{eqnarray}
\label{part8}
\Xi(T,L,D,\mu)\,& = & \, \exp\left[-\frac{\Omega(T,L,D,\mu)}{k_{B}T}\right] \,\\
& = &\,    -i\,\exp\left[ \frac{\beta L^2D}{2a}\,\mu^2\right]\,\sqrt{\frac{L^2D}{2\pi \beta a}}\,\int_{\beta \alpha - i\infty}^{\beta \alpha + i\infty} ds 
\,\exp(-L^2D\,\varphi(s))\nonumber \quad,
\end{eqnarray}
where 
\begin{eqnarray}
\label{part9}
&\varphi(s) &\, = \, \varphi_{b}(s) -\frac{1}{\lambda^3}\,\left\{\frac{\lambda}{D}\,\sum_{k_{z}}\,
g_{2}\left[ \exp(s-\beta\epsilon(k_z) ) \right] \,-\,g_{5/2}[\exp(s)] \right\} \\
&+& \,\frac{1}{L^2D}\,\sum_{k_{z}} \left\{\ln\left[1- \exp\left(s-\beta\epsilon(k_z) \right) \right] \,
 - \,\ln[1-\exp(s)] \right\} \nonumber 
\end{eqnarray}
Here $\epsilon(k_z) = \hbar^{2}k_{z}^2/2m$, and $k_{z}=2\pi n/D, \, n=0, \pm1, \pm2,\cdots $ due to periodic boundary conditions.
Equation (\ref{part9}) follows directly from the equality
\begin{eqnarray}
\label{rownosc}
\ln\Xi_{0} (T,L,D,s/\beta) \,= - \sum_{k_{z} \ne 0}\,\ln\{ 1 - \exp[s-\beta\epsilon(k_z)]\} \, - 
\,\sum_{k_{z}}\left(\frac{L}{\lambda}\right)^{2}\,g_{2}[ \exp[s-\beta\epsilon(k_z) ] 
\end{eqnarray}
valid  for $L/\lambda\gg 1$. The first term on the rhs of (\ref{rownosc}) corresponds to the $k_{x}=k_{y}=0$ contribution to the 
series defining the logarithm of the perfect gas partition function
\be
\label{ksizero}
\ln\Xi_{0} (T,L,D,s/\beta) = - \sum_{\bk}\,\ln [1-\exp(s - \beta\epsilon(\bk))] \quad.
\ee 
Finally, the bulk density $\varphi_{b}(s)$ which we added and subtracted when writing (\ref{part9}) 
has been defined in Eq.(\ref{part6}).

Taking in Eq.(\ref{part9}) the $L\to\infty $  limit we obtain the formula
\be
\label{part99}
\varphi(s) \, =  \, \varphi_{b}(s) -\frac{1}{\lambda^3}\,\left\{\frac{\lambda}{D}\,\sum_{k_{z}}\,
g_{2}\left[ \exp[s-\beta\epsilon(k_z) ] \right] \,-\,g_{5/2}[\exp(s)] \right\} 
\ee
corresponding to infinite walls. With the help of the Jacobi identity \cite{MZ2006, ZUK77}
\begin{eqnarray}
\sum_{n=-\infty}^{\infty} \, \exp( - \pi \kappa n^2) \, = \, \frac{1}{\sqrt{\kappa}} \, \sum_{n=-\infty}^{\infty} \, \exp(- \pi n^2 /\kappa ) 
\end{eqnarray}
valid for $\kappa > 0$, and relation 
\be
\label{zwiazek}
\beta\epsilon(k_z) = \left( \frac{\lambda}{D} \right)^{2}\,\pi\,n^2 \quad,
\ee
Eq.(\ref{part99}) can be rewritten in the following form
\begin{eqnarray}
\label{part90}
\varphi(s) \,=\, \varphi_{b}(s) -\frac{2}{\lambda^3}\,\sum_{r=1}^{\infty} \, \sum_{n=1}^{\infty}  \frac{1}{r^{\frac{5}{2}}}
\exp\left(sr -\frac{ D^2\,\pi \,n^2}{\lambda^2 r}\right) \quad.
\end{eqnarray}

The saddle point $\bar{s}$ of function $\varphi(s)$ satisfies the equation
\begin{eqnarray}
\label{part91}
0\, = \varphi'(\bar{s}) \,=\, \varphi'_{b}(\bar{s})   -\frac{2}{\lambda^3}\,\sum_{r=1}^{\infty} \, \sum_{n=1}^{\infty}  
\frac{1}{r^{\frac{3}{2}}} \exp\left(\bar{s}r -\frac{ D^2\,\pi \,n^2}{\lambda^2 r}\right) \quad.
\end{eqnarray}
In fact, we are interested in the behavior of $\bar{s}$ for large values of the dimensionless width $D/\lambda$, and the sole
property of $\bar{s}$ we will need is the fact that 
in the limit $D/\lambda\to\infty $  the point $\bar{s}$ approaches the bulk saddle point $s_{0}$.
Indeed, within the steepest descent method the solution $\bar{s}$ is inserted into $\varphi(s)$ 
in order to evaluate the surface contribution (\ref{root3}) to the grand canonical potential.  
In the case of periodic boundary conditions the excess grand canonical free energy density $\omega_{s}(T,D,\mu)$ 
represents exclusively the fluctuation-induced interaction between the walls; there is no contribution to $\omega_{s}(T,D,\mu)$ 
stemming from two noninteracting  wall-boson gas interfaces. \\
The above remarks combined with Eq.(\ref{part90}) imply that for large $D/\lambda$ the dominant contribution to $\omega_{s}(T,D,\mu)$ 
has the form
\begin{eqnarray}
\label{surf1}
\omega_{s}(T,\mu,D) \,=\, -\frac{2\,D\,k_{B}T}{\lambda^3}\,\sum_{r=1}^{\infty} \, \sum_{n=1}^{\infty}  
\frac{1}{r^{\frac{5}{2}} }\exp\left[ s_{0}r - \frac{D^2\,\pi \,n^2}{\lambda^2 r}\right] \quad,
\end{eqnarray}
where $s_{0}$ is the bulk saddle point defined by Eq.(\ref{root1}). From the previous analysis we know that for $\mu/an_c < 1$, $s_0$ 
takes a negative value ensuring the convergence of the series in Eq.(\ref{surf1}).

For $D/\lambda \gg1$ the sum over $r$ on the rhs of Eq.(\ref{surf1}) can be transformed as \cite{MZ2006, AS1972}  
\begin{eqnarray}
\label{sum1}
\sum_{r=1}^{\infty} \, 
\frac{1}{r^{\frac{5}{2}} }\exp\left[ s_{0}r - \frac{D^2\,\pi \,n^2}{\lambda^2 r}\right] \,\simeq \,
\int_{0}^{\infty} d\rho \frac{1}{\rho^{\frac{5}{2}} }\exp\left[ s_{0}\rho - \frac{D^2\,\pi \,n^2}{\lambda^2 \rho}\right]\,  \noindent 
\end{eqnarray}
\[ = \frac{1}{2 \pi}\,\frac{\lambda^3}{D^3}\,\frac{1}{n^3}\,\left(1 \,+\, 2n \frac{D}{\kappa_{per}}\right)\,
\exp\left(- 2n \frac{D}{\kappa_{per}}\right) \quad, \]
where
\begin{eqnarray}
\label{cor1}
\kappa_{per} \, =\, \frac{\lambda}{\sqrt{\pi |s_{0}|}} 
\end{eqnarray}
is the characteristic decay length. 

We have already noticed that $s_{0}\to 0$ when  the condensation point $\mu \,=\,a n_{c}$ is approached from below. 
It is instructive to determine the structure of the the solution of Eq.(\ref{root2}) in the vicinity of the condensation point. 
With the help of the asymptotic relation \cite{ZUK77} 
\begin{eqnarray}
\label{cor2}
g_{3/2}\left(\exp(s_{0})\right)\,\approx \,-2\sqrt{\pi} \,(-s_{0})^{1/2}\,+\zeta(3/2) 
\end{eqnarray}
we find from Eq.(\ref{root2}) that 
\begin{eqnarray}
\label{cor3}
|s_{0}|^{1/2}\,=\,\frac{(a n_{c}-\mu )}{an_{c}}\,\frac{\zeta(3/2)}{2 \pi^{1/2}} \quad.
\end{eqnarray}
Near the condensation point the behavior of the decay length (\ref{cor1}) is thus described by 
\begin{eqnarray}
\label{cor4}
\kappa_{per} \, =\, \lambda \,\frac{an_{c}}{(a n_{c}-\mu )}\,\frac{2 \pi^{1/2}}{\zeta(3/2)}  \quad.
\end{eqnarray} 
Thus upon approaching the condensation point from the one-phase region the decay length $\kappa_{per}$ diverges like $ (a n_{c}-\mu)^{-1}$. 
We notice that the value of the relevant critical exponent $\nu_{IMP}=1$ 
differs from that corresponding to the perfect Bose gas, where
$\kappa_{per} \, \sim \, (-\mu)^{-1/2}$ with the corresponding critical exponent $\nu_{P}=1/2$ (see \cite{MZ2006}). 

For the imperfect Bose gas the Casimir force 
\begin{eqnarray}
F(T,D,\mu)=-\frac{\partial \omega_{s}^{<}(T,D,\mu)}{\partial D} = 
 k_{B}T \, \frac{\partial }{\partial D} \left\{ \frac{1}{\pi D^2}  \sum_{n=1}^{\infty}  \frac{1 + 2n D/\kappa_{per} }{n^3} \, 
\exp\left[- 2n \frac{D}{\kappa_{per}}\right] \right\}
\end{eqnarray}
can be rewritten with the help of the scaling function \cite{GD2006} 
\begin{eqnarray}
\Psi(x) \,=\,-\sum_{n=1}^{\infty} \, \frac{1 \,+\, 2n x}{\pi n^3} \, \exp(- 2n x) \quad,
\end{eqnarray} 
such that $\Psi(0)=-\zeta(3)/\pi$, $\Psi'(0)= 0$, and $\Psi(x \gg 1)\,\approx -2/\pi \,x\,e^{-2x}$. Putting $x=D/\kappa_{per}$, 
we find in the one-phase region ({\it normal phase}) near the condensation point 
\begin{eqnarray}
\label{explaw}
\frac{F(T,D,\mu)}{k_{B}T}\,=\,\,\frac{1}{ D^3} \,\left[ 2\,\Psi(x)\,-\,x\,\Psi'(x)\,\right] \quad. 
\end{eqnarray} 
We note that the scaling function $2\,\Psi(x)\,-\,x\,\Psi'(x)$ in Eq.(\ref{explaw}) depends on the ratio $x=D/\kappa_{per}$ of the width 
$D$ and the decay length $\kappa_{per}$, 
see Eq.(\ref{cor1}). An interesting problem is to determine the correlation length $\xi$ characterizing the decay of the density-density 
correlation function of the 
imperfect Bose gas and then to express the variable $x$ as function of $D/\xi$. This would allow one to calculate the universal scaling 
function for the Casimir force \cite{KD1992, GD2006}. 

On the other hand, in the two-phase region (in the presence of condensate) one observes the power-law decay 
\begin{eqnarray}
\label{powerlaw}
\frac{F(T,D, \mu > an_{c})}{k_{B}T}\,=\, -\frac{2 \zeta(3)}{\pi}\,\frac{1}{D^3} \quad,
\end{eqnarray}
which is exactly the same, and with the same amplitude (universal when divided by $k_{B}T$) as in the perfect Bose gas \cite{MZ2006} 
at $\mu=0$. For comparisons with analytical and numerical results for the Casimir amplitude and the scaling function    
for the XY universality class which have been obtained via various methods, inter alia via Monte Carlo simulations and the 
field-theoretical renormalization-group theory for $O(N)$ symmetric systems, see  
\cite{KD1992, MAC2007, ZAN2007, VAS2007, VAS2009, HAS2010, HUC2007}. \\

The above analysis can be straightforwardly extended to the case of Dirichlet and Neumann boundary conditions. 
We thus present here only the final results. In these cases - contrary to the case of periodic boundary conditions 
- one obtains nonzero values for the single wall-Bose gas surface free energy density $\sigma(T,\mu)$. 
The excess grand canonical free energy density is the sum of $2\sigma(T,\mu)$, and the contribution
due to effective interaction between the walls. 

For $\sigma^{<}(T,\mu)$ one finds 
\begin{eqnarray}
\label{DN1}
\sigma^{<}(T,\mu) \, = \,\pm  \frac{k_{B}T}{4 \lambda^2}\,g_{2}[\exp(s_{0})] \quad, 
\end{eqnarray} 
where $s_{0}$ is the solution of Eq.(\ref{root2}), and the $(+)$ and $(-)$ signs correspond to the Dirichlet and Neumann boundary conditions, respectively. 
The expressions (\ref{DN1}) for the surface free energy density should be compared with the 
corresponding results obtained for the perfect Bose gas, see Chapter 5 in \cite{ZUK77}. One concludes that although in general these expressions are different, 
they become identical when evaluated  in the two-phase region, i.e., 
$\mu=0$ for the perfect gas, and $\mu \ge an_{c}$ for the imperfect gas.     
 
The contribution to the excess grand canonical free energy density due to the effective interaction between the walls is given by
\begin{eqnarray}
-\frac{k_{B}T}{8\,\pi D^2} \, \sum_{n=1}^{\infty} \, \frac{1 \,+\, 2n D/\kappa_{D,N} }{n^3} \, \exp(- 2n D/\kappa_{D,N} ) \quad,
\end{eqnarray}
where the expression for the decay length $\kappa_{D,N} $ is the same in the Dirichlet and Neumann  cases 
\begin{eqnarray}
\label{cor5}
\kappa_{D,N} \, =\, \frac{\lambda}{2 \sqrt{\pi |s_{0}|}} \quad ,
\end{eqnarray}
and differes by factor 2 from the corresponding expression $\kappa_{per}$ in the periodic case (see Eq.(\ref{cor1})).

In the presence of condensed phase, where $\mu \ge \mu_{c}=an_{c}$, one observes both in the Dirichlet and in the Neumann case 
the same power law decay of the corresponding Casimir force 
\begin{eqnarray}
\frac{F(T,D,\mu\ge a n_{c})}{k_{B}T}\,=\, -\frac{\zeta(3)}{4 \pi}\,\frac{1}{D^3} \quad, 
\end{eqnarray}
with the same amplitude (universal when divided by $k_{B}T$) as in the perfect Bose gas for these boundary conditions (see \cite{MZ2006}).  

\section{Concluding comments}

The aim of our study was to investigate Casimir forces induced by fluctuations of an imperfect Bose gas contained between two infinite 
parallel plane walls situated at distance $D$, and compare the results with those obtained previously 
in Ref.\cite{MZ2006} for a perfect gas.
We based our approach on exact relation (\ref{part4}) between the grand canonical partition function of an imperfect (mean-field)
gas, and that of an perfect gas. The derivation of Eq.(\ref{part4}) was based on the fact that the
mean-field Hamiltonian, Eq.(\ref{HMF}), allowed 
the use of the occupation number representation (for a systematic description of the
so-called diagonal models see the review \cite{ZB2001}).

We first applied to Eq.(\ref{part4}) the steepest descent method to determine the bulk free energy density.
In this way we obtained the mean-field equation of state, previously established rigorously by other methods.
It might be interesting to note that the self-consistent relation (\ref{free11}) has been derived and studied 
for an interacting Bose gas in a much more general context \cite{MP2003}.

The evaluation of the excess grand canonical free energy density under periodic boundary conditions, presented in Section III, 
used in an essential way the fact that the saddle point corresponding to finite distance $D$ approched the 
bulk saddle point when $D\to\infty$. 
Our final results,  Eqs (\ref{explaw}, \ref{powerlaw}) confirmed the appearance of power law decay $(\sim D^{-3})$ of the Casimir force in the presence 
of condensate ($\mu \ge an_{c}$), and of exponential decay in the one phase region ($\mu < an_{c}$). Here our predictions reproduced
exactly the situation in a perfect Bose gas. 
However, we noticed a difference in the value of exponent $\nu$ 
governing the divergence of the characteristic length scale at the approach from below to the critical value of the 
chemical potential, $\mu_c = an_c$. In a perfect gas $\nu_{P} = 1/2$, whereas in the presence of  the repulsive mean-field
$\nu_{IMP} = 1$, as shown in Eq.(\ref{cor4}). This difference seems to be the only manifestation of the mean-field coupling between the
bosons as far as the Casimir effect is concerned.

Finally, let us make the following comment. Because of  non-equivalence of the grand canonical and canonical
ensembles of a perfect gas one could wonder how the results derived in Ref.\cite{MZ2006} would compare
with calculations  based on the canonical free energy. It turns out that the answer to this important question can
be readily derived from the remarkable, rigorous analysis of the properties of a perfect Bose gas in a finite volume given
in Ref.\cite{ZUK77}, see especially Section 3. We checked that as far as the Casimir force is concerned 
both grand canonical and canonical ensemble give exactly the same results. The anomalies of the grand canonical 
fluctuations in the total occupation number of the ground state in the presence of condensate have thus no effect 
on the large distance behavior of the Casimir force.

\end{document}